\documentclass[12pt,journal,draftclsnofoot,journalpaper,onecolumn]{IEEEtran} %TWC
\usepackage{amsmath}
\usepackage{amsfonts}
\usepackage{mathrsfs}
\usepackage{cite}
\usepackage{graphicx}
\usepackage{amssymb}
\usepackage{stfloats}
\usepackage{subeqnarray}%(1a)
\usepackage{cases}
\usepackage{indentfirst}

\begin{document}

\title{Relay Selection for Bidirectional AF Relay Network with Outdated CSI}

\author{\IEEEauthorblockN{Hongyu Cui, Rongqing Zhang, Lingyang Song, and Bingli
Jiao}\\
\IEEEauthorblockA{School of Electronics Engineering and Computer
Science\\ Peking University, Beijing, China, $100871$\\
%Email: \{\}@pku.edu.cn.\\
}}

\maketitle
\begin{abstract}
Most previous researches on bidirectional relay selection~(RS)
typically assume perfect channel state information~(CSI). However,
outdated CSI, caused by the the time-variation of channel, cannot be
ignored in the practical system, and it will deteriorate the
performance. In this paper, the effect of outdated CSI on the
performance of bidirectional amplify-and-forward RS is investigated.
The optimal single RS scheme in minimizing the symbol error
rate~(SER) is revised by incorporating the outdated channels. The
analytical expressions of end-to-end signal to noise ratio~(SNR) and
symbol error rate~(SER) are derived in a closed-form, along with the
asymptotic SER expression in high SNR. All the analytical
expressions are verified by the Monte-Carlo simulations. The
analytical and the simulation results reveal that once CSI is
outdated, the diversity order degrades to one from full diversity.
Furthermore, a multiple RS scheme is proposed and verified that this
scheme is a feasible solution to compensate the diversity loss
caused by outdated CSI.
\end{abstract}


\newpage
\section{Introduction}

Recently, bidirectional relay communications, in which two sources
exchange information through the intermediate relays, have attracted
a lot of attention, and different transmission schemes of
bidirectional relay have been proposed
in\cite{Shengli2009,Louie2010,Popovski2007}. An
amplify-and-forward~(AF) based network coding scheme, named as
analog network coding~(ANC), was introduced in \cite{Popovski2007}.
With ANC, the data transmission of bidirectional AF relay can be
divided into two phases, and the spectral efficiency can get
improved \cite{Popovski2007}. Recently, relay selection~(RS)
technique for bidirectional relay networks has been intensively
researched, due to its ability to achieve full diversity with only
one relay\cite{Jing2009,Upadhyay2011,Song2011,Kyu2009,Nguyen2010}.
Performing RS, the best relay is firstly selected before data
transmission, according to the predefined RS scheme. In
\cite{Song2011}, a optimal RS scheme in minimizing the average
symbol error rate~(SER) for the source pair was proposed, and the
bounds of SER and the optimal power allocation scheme were provided.
The author in \cite{Jing2009} derived the tight lower bound of block
error rate for the bidirectional RS network. The performance bounds,
such as the average sum rate and outage probability, for the
bidirectional RS was offered under the Rayleigh fading in
\cite{Kyu2009}, and these bounds were extended to the Nakagami-m
fading in \cite{Upadhyay2011}. In \cite{Liu2010}, a relay-assisted
bidirectional cellular network was considered, and a resource
allocation method, including the optimal relay selection scheme, was
proposed to improve the overall system performance. The diversity
order for various RS schemes of bidirectional RS was studied in
\cite{Nguyen2010}, and it proved that the RS schemes can achieve
full diversity when the channel state information~(CSI) is perfect.

Furthermore, all the aforementioned researches analyzed the
bidirectional RS with perfect CSI. Outdated CSI, caused by the
time-variation of channel, cannot be negligible in the practical
system, and it makes the selected relay not the best for the data
transmission. The impact of outdated CSI has been fully discussed in
one-way RS\cite{Michalopoulos2012,Soysa2012,Seyi2011,Dong2012}. In
\cite{Michalopoulos2012,Soysa2012}, the expressions of SER and
outage probability for one-way AF RS were obtained, and the partial
RS and opportunistic RS were both considered with outdated CSI. The
impact of outdated CSI and channel estimation error on the one-way
decode-and-forward~(DF) RS was analyzed in \cite{Seyi2011}. Multiple
RS with AF and DF protocols was considered in one-way relay with
outdated CSI \cite{Dong2012}, in which the outage probability and
diversity order were analyzed. In \cite{Fan2012}, the two-way
network with one relay and multiple users was studied, and the
effect of outdated CSI on user selection was researched. In
\cite{Amarasuriya2012}, the antenna selection criterion of MIMO
two-way relay was proposed, and the performance with outdated CSI
was analyzed when there are one single-antenna relay.

However, to the best of the authors' knowledge, the impact of
outdated CSI on the performance of bidirectional RS has not been
investigated. In this paper, we analyze the SER performance of the
bidirectional AF RS with outdated CSI. The optimal single RS in
minimizing the instantaneous SER is revised by incorporating the
outdated channels. The distribution of end-to-end signal-to-noise
ratio~(SNR), the analytical average SER expressions are derived in
this paper, and verified by the Monte-Carlo simulations. The effect
of the parameters, such as the number of relays and the correlation
coefficient of outdated CSI, are investigated. The theoretical
analysis and the simulation results reveal that once CSI is
outdated, the diversity order reduces to one, regardless of the
number of relays. Furthermore, a multiple RS scheme for the
bidirectional relay is proposed to improve the diversity loss.

In summary, the main contribution of this paper is listed as
follows:
\begin{enumerate}
  \item Outdated CSI is taken into account to derive the analytical results of bidirectional RS, and its therein impact is
  investigated.
  \item Considering the generalized network structure, i.e., different
channels have different variances and different correlation
coefficients of outdated CSI, the generalized average SER expression
is obtained, which can be further simplified according to the
concrete situations, such as high SNR analysis and the analysis of
symmetric network.
\item The SER expression derived in this paper is
tight with the exact result, and verified by the Monte-Carlo
simulations.
\item A multiple RS scheme by selecting the best $K$ relays from $N$
available relays is proposed in the bidirectional relay, and the
diversity order is analyzed, which reveals that the multiple RS can
compensate the diversity loss caused by outdated CSI.
\end{enumerate}

The remainder of this paper is organized as follows: In Section
\uppercase\expandafter{\romannumeral2}, the system model of
bidirectional AF RS, the outdated CSI model, and the RS schemes are
described in detail. Section \uppercase\expandafter{\romannumeral3}
provides the analytical expressions of bidirectional RS, including
the distribution function of received SNR, the performance of
end-to-end average SER, and the diversity order. Simulation results
and performance analysis are presented in Section
\uppercase\expandafter{\romannumeral4}. Finally, Section
\uppercase\expandafter{\romannumeral5} concludes this paper.

\emph{Notation:}~$\left|\cdot\right|$ represents the absolute value,
$\mathbb{E}$ is used for the expectation, and $\Pr$ represents the
probability. The probability density function~(PDF) and the
cumulative probability function~(CDF) of random variable~(RV) $x$
are denoted by $f_x\left(\cdot\right)$ and $F_x\left(\cdot\right)$,
respectively.

\section{System Model}

\begin{figure}[h!]
\centering
\includegraphics[height=2.8in,width=6.2in]{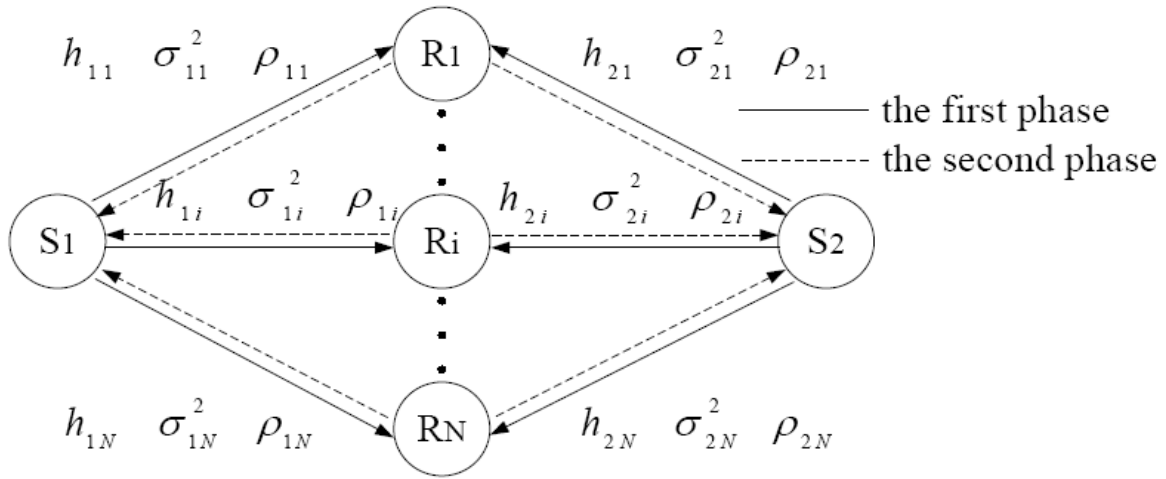}
\caption{System model of bidirectional relay network.}
\label{fig:SM}
\end{figure}

As shown in Fig.~\ref{fig:SM}, the system investigated in this paper
is a bidirectional AF relay network with two sources $S_j$, $j=1,2$,
exchanging information through $N$ relays $R_i$, $i=1,\ldots,N$, in
which each communication node is equipped with a single half-duplex
antenna. The transmit powers of each source and each relay are
denoted by $p_s$ and $p_r$, respectively. The direct link between
the sources does not exist due to the shadowing effect, and the
channel coefficients between $S_j$ and $R_i$ are reciprocal, denoted
by $h_{ji}$. All the channel coefficients follow independent
complex-Gaussian distribution with zero mean and variance of
$\sigma_{ji}^2$.

\subsection{Instantaneous Received SNR at the Sources}

Considering the transmission via $R_i$, the data transmission of
bidirectional AF relay is divided into two phases. During the first
phase, the sources simultaneously send their respective information
to $R_i$. The received signal at $R_{i}$ is
$r_{i}=\sqrt{p_s}h_{1i}s_1+\sqrt{p_s}h_{2i}s_2+n_{ri}$, where $s_j$
denotes the modulated symbols transmitted by $S_j$ with the average
power normalized, and $n_{ri}$ is the additive white Gaussian
noise~(AWGN) at $R_i$, with zero mean and variance of $\sigma_n^2$.
During the second phase, $R_i$ amplifies the received signal and
forwards it back to the sources. The signal generated by $R_i$
satisfies $t_i = \sqrt {p_r } \beta_i r_i$, where $\beta_i =
\left(p_{s} |h_{1i}|^2+ p_{s} |h_{2i}|^2 + \sigma_n^2 \right)^{
-1/2} $ is the variable-gain factor\cite{Song2011}. The received
signal at $S_j$, $j=1,2$, is $y_{j} = h_{ji}t_i + n_{sj}$, where
$n_{sj}$ is the AWGN at $S_j$. Then, after canceling the
self-interference, i.e., $\sqrt{p_s p_r} \beta_i h_{ji} h_{ji}s_j$,
the instantaneous received SNR at $S_j$ via $R_i$ is\cite{Song2011}
\begin{equation}\label{Eq:gamma}
\gamma _{ji}  = \frac{{\psi _s \psi _r \left| {h_{ji} } \right|^2
\left| {h_{\bar ji} } \right|^2 }}{{\left( {\psi _s  + \psi _r }
\right)\left| {h_{ji} } \right|^2  + \psi _s \left| {h_{\bar ji} }
\right|^2  + 1}}
\end{equation}
where $\psi_s =p_s/ \sigma_n^2$, $\psi_r=p_r/\sigma_n^2$, and
$\{j,\overline j\}=\{1,2\}~\mbox{or}~\{2,1\}$.

Furthermore, by ignoring the constant $1$ in the denominator of
\eqref{Eq:gamma}, we can obtain the upper bound of received SNR,
i.e.,
\begin{align}\label{Eq:gammahigh}
\gamma _{ji}  = \frac{{\left( {\psi _r \left| {h_{ji} } \right|^2 }
\right)\left( {\psi _h \left| {h_{\bar ji} } \right|^2 }
\right)}}{{\left( {\psi _r \left| {h_{ji} } \right|^2 } \right) +
\left( {\psi _h \left| {h_{\bar ji} } \right|^2 } \right)}}
\end{align}
where $\psi_h=\psi_s\psi_r/\left(\psi_s+\psi_r\right)$. This bound
is tight enough with the exact result, especially in high SNR.
Therefore, in the following, we use the bound for analysis
\cite{Song2011}.

\subsection{Relay Selection Schemes}

To minimize the instantaneous SER for the source pair, the index of
the selected relay should satisfy\cite{Song2011}
\begin{align}\label{Eq:optimal1}
k=\arg \max_{i} \min \left\{\gamma_{1i},\gamma_{2i}\right\}
\end{align}
where $\gamma_{ji}$ is decided by \eqref{Eq:gammahigh}.

Before further discussion and analysis, we provide the following
Lemma.

\emph{\textbf{Lemma 1:}} The minimization of $\gamma_{1i}$ and
$\gamma_{2i}$ is bounded by
\begin{align}\label{Eq:Lemma1}
\min \left(\gamma_{1i},\gamma_{2i}\right) \le
\frac{\psi_s\psi_r}{\psi_s+\psi_r}\min
\left(|h_{1i}|^2,|h_{2i}|^2\right)
\end{align}
where the right-hand side of \eqref{Eq:Lemma1} is the upper bound of
the left-hand side, and it is also a tight approximation, especially
in high SNR.

\emph{\textbf{Proof: }}The derivation is given in Appendix A.$\hfill
\blacksquare$

According to Lemma~1 and \eqref{Eq:optimal1}, the optimal RS scheme
in minimizing the instantaneous SER for the source pair is
equivalent to \cite{Jing2009,Upadhyay2011}
\begin{align}\label{Eq:perfectRS}
k=\arg \max_i \min \left\{|h_{1i}|^2,|h_{2i}|^2\right\}.
\end{align}

Specifically, the relay selection can be achieved in the distributed
or centralized manner.

If the relay selection is conducted in the distributed manner
\cite{Bletsas2006, Li2011}, each relay $R_{i}$ estimates the local
channel coefficients $h_{1i}$ and $h_{2i}$, by the exchanges of
control packets, such as, ready-to-send and clear-to-send frames
\cite{Bletsas2006}. The concrete estimation method can be found in
\cite{Gao2009}, which is beyond the scope of this paper. In the
selection process, the timer mechanism is employed among the
available relays to determine the ``best'' relay autonomously
\cite{Bletsas2006}. In this procedure, the delay between relay
selection and data transmission takes up about one cooperation phase
\cite{Li2011}, which may subject to relatively serious channel
variations, and thus the CSI is outdated.

If the relay selection is conducted in the centralized manner
\cite{Michalopoulos2012}, the central unit, such as, the source
$S_1$, estimates all the links' channel coefficients, with the help
of the pilots from the other source $S_2$. The concrete estimation
method can be found in \cite{Michalopoulos2012}. Based on the
estimated channel coefficients, the ``best" relay is selected,
according to the predefined RS schemes. Then, the central unit
broadcasts the index of the selected relay to all the relays. In
this procedure, the delay between relay selection and data
transmission also exists, due to the feedback delay
\cite{Michalopoulos2012}, and thus the CSI is also outdated.

In summary, because of the feedback delay and the scheduling delay,
the selection of the best relay is not based on the current time
instant\cite{Michalopoulos2012}, regardless of centralized and
distributed relay selection. The channel coefficient at the
selection instant is denoted by $\hat h_{ji}$. Due to the
time-variation of channel, $\hat h_{ji}$ is outdated to $h_{ji}$,
and their relationship is decided by the Jakes'
model\cite{Michalopoulos2012}
\begin{align}\label{Eq:hf}
\hat h_{ji}=\rho_{ji}h_{ji}+\sqrt{1-{\rho_{ji}^2}}\varepsilon_{ji}
\end{align}
where $\varepsilon_{ji}$ is an independent identically distributed
RV with $h_{ji}$, the correlation coefficient
$\rho_{ji}=J_0\left(2\pi f_{d_{ji}}T_d\right)$, where $J_0\left(
\cdot \right)$ stands for the zeroth order Bessel
function\cite{Abramowitz}, $f_{d_{ji}}$ is the Doppler spread, and
$T_d$ is the time delay between $\hat h_{ji}$ and $h_{ji}$.
Moreover, $\rho_{{ji}}=1$, i.e., $f_{d_{ji}}=0$, means CSI is
perfect, and $\rho_{{ji}}<1$, i.e., $f_{d_{ji}}>0$, means CSI is
outdated.

Therefore, the RS scheme \eqref{Eq:perfectRS} with outdated CSI is
converted into
\begin{align}\label{Eq:outdatedRS}
k=\arg \max_i \min \left\{|\hat h_{1i}|^2,|\hat h_{2i}|^2\right\}.
\end{align}

In the following, we analyze the performance of the RS scheme
\eqref{Eq:outdatedRS} with outdated CSI, and the performance with
perfect CSI can be obtained by setting $\rho_{ji}=1$.

\section{Performance Analysis of Bidirectional Relay Selection with Outdated CSI}

In the following, the analytical and asymptotic average SER
expressions of the single RS scheme \eqref{Eq:outdatedRS} are
derived in a closed-form.

\subsection{The Distribution of end-to-end Received SNR}

To analyze the performance of bidirectional AF single RS
\eqref{Eq:outdatedRS}, the distribution function of $\gamma_{jk}$ in
\eqref{Eq:gammahigh} is required. Therefore, the analytical PDF and
CDF of \eqref{Eq:gammahigh} with outdated CSI are derived in this
part.

In order to obtain the exact distribution of \eqref{Eq:gammahigh},
we need to achieve the distribution of $ \left| { h_{j{k}} }
\right|^2$, which is decided by $\left| \hat  h_{jk}  \right|^2$,
according to \eqref{Eq:hf}. Moreover, $\left| \hat  h_{jk}
\right|^2$ is decided by the RS scheme \eqref{Eq:outdatedRS}. After
some manipulation, we can obtain

\textbf{\emph{Lemma 2: }} The PDF of $\left| {h_{jk} } \right|^2 $,
$\left\{j,\overline{j}\right\}=\left\{1,2\right\}~\mbox{or}~\left\{1,2\right\}$
is
\begin{align}\label{Eq:lemma21}
f_{\left| {h_{jk} } \right|^2 } \left( z \right) = \sum\limits_{i =
1}^N {\sum\limits_{t = 0}^{N - 1} {\sum\limits_{A_t } {\left( { - 1}
\right)^t \left( {1 + \sum\limits_{l \in A_t } {\frac{{\sigma _{\bar
ji}^2 }}{{\sigma _l^2 }}} } \right)} } ^{ - 1} }
\bigg[\frac{1}{{\sigma _{ji}^2 }}\exp \left( { - \frac{z}{{\sigma
_{ji}^2 }}} \right) + \frac{{\zeta _j }}{{\sigma _{ji}^2 }}\exp
\left( { - \frac{{\xi _j }}{{\sigma _{ji}^2 }}z} \right)\bigg]
\end{align}
where
\begin{align}\label{Eq:xi}
\xi _j  = \left( {\frac{1}{{\sigma _i^2 }} + \sum\limits_{l \in A_t
} {\frac{1}{{\sigma _l^2 }}} } \right)\left( {\frac{{\rho _{ji}^2
}}{{\sigma _{ji}^2 }} + \frac{{1 - \rho _{ji}^2 }}{{\sigma _i^2 }} +
\sum\limits_{l \in A_t } {\frac{{1 - \rho _{ji}^2 }}{{\sigma _l^2
}}} } \right)^{ - 1} ,
\end{align}
\begin{align}\label{Eq:zeta}
\zeta _j  = \left( {\frac{{\sigma _{\bar ji}^2 }}{{\sigma _{ji}^2
}}\sum\limits_{l \in A_t } {\frac{1}{{\sigma _l^2 }}} }
\right)\left( {\frac{{\rho _{ji}^2 }}{{\sigma _{ji}^2 }} + \frac{{1
- \rho _{ji}^2 }}{{\sigma _i^2 }} + \sum\limits_{l \in A_t }
{\frac{{1 - \rho _{ji}^2 }}{{\sigma _l^2 }}} } \right)^{ - 1} .
\end{align}

In addition, $\sum\limits_{A_t }$ is the abbreviation of
$\sum\limits_{\scriptstyle A_t \subseteq \left\{ {1, \ldots ,N}
\right\}\backslash i \hfill \atop
  \scriptstyle \left| {A_t } \right| = t \hfill}$, $\left| {A_t } \right|$ represents the
cardinality of set $A_t$, and
$\sigma_{i}^2=\sigma_{1i}^2\sigma_{2i}^2/\big(\sigma_{1i}^2+\sigma_{2i}^2\big)$
$i=1,\ldots,N$.

\emph{\textbf{Proof: }} The derivation is given in Appendix B.
$\hfill \blacksquare$

Before deriving the distribution of received SNR, we introduce two
equations, which are necessary for the following analysis.

\textbf{\emph{Lemma 3: }}
\begin{align}\label{Eq:lemma31}
\sum\limits_{i = 1}^N {\sum\limits_{t = 0}^{N - 1} {\sum\limits_{A_t
} {\left( { - 1} \right)^t \left( {\sum\limits_{l \in A_t }
{\frac{{\sigma _{i}^2 }}{{\sigma _{l}^2 }} + 1} } \right)^{ - 1} } }
}  = 1, \mbox{and} \sum\limits_{t = 1}^{N - 1} {\sum\limits_{A_t }
{\left( { - 1} \right)^t \left( {\sum\limits_{l \in A_t }
{\frac{1}{{\sigma _{l}^2 }}} } \right)^k } }  = 0,0 \le k \le N - 2.
\end{align}

\emph{\textbf{Proof: }}The derivation is given in Appendix C.
$\hfill \blacksquare$

According to Lemma 3, we can obtain the CDF of $ |h_{jk}|^2$,
$j=1,2$, by integrating the PDF in Lemma~2. Also, the PDF and CDF of
$\Omega_1=\psi_r|h_{jk}|^2$ and $\Omega_2=\psi_h
|h_{\overline{j}k}|^2$ in \eqref{Eq:gammahigh} can be obtained by
the fact that when $ Y = mX \left(m> 0\right) $, $f_Y \left(z\right)
= \left({1}/{m}\right)f_X \left({z}/{m}\right)$ and $F_Y \left( z
\right) = F_X \left( {z}/{m} \right) $\cite{Paoulis}.

\textbf{\emph{Proposition 1:}} With the definition that
\begin{align}\label{Eq:abb}
a_{i}=\frac{1}{{ \psi_r \sigma _{ji}^2 }}~,~ b_{i'}=\frac{1}{{
\psi_h \sigma _{\bar j i'}^2 }},
\end{align}
the CDF of the received SNR at $S_j$ via the selected relay $R_k$ is
\begin{align}\label{Eq:gamma1cdf}
&F_{\gamma _{jk} } \left( z \right) = 1 - \sum\limits_{i = 1}^N {\sum\limits_{t = 0}^{N - 1} {\sum\limits_{A_t } {\sum\limits_{i' = 1}^N {\sum\limits_{t' = 0}^{N - 1} {\sum\limits_{A_{t'} } {\left( { - 1} \right)^{t + t'} \left( {1\hspace{-0.8mm}  +\hspace{-0.8mm}  \sum\limits_{l \in A_t } {\frac{{\sigma _{\bar ji}^2 }}{{\sigma _{l}^2 }}} } \right)^{ - 1} \left( {1 \hspace{-0.8mm} +\hspace{-0.8mm}  \sum\limits_{l' \in A_{t'} } {\frac{{\sigma _{ji'}^2 }}{{\sigma _{l'}^2 }}} } \right)^{ - 1} \sqrt {4a_{i} b_{i'} } } } } } } }  \notag\\
&\hspace{64mm}\times \left( {f_{11}  + f_{12}  + f_{21}  + f_{22} }
\right)
\end{align}
where
\begin{align}
f_{11}  = z\exp \left( { - a_{i} z -  b_{i'} z} \right)K_1 \left(
{2z\sqrt { a_{i} b_{i'} } } \right),
\end{align}
\begin{align}
f_{12}  = \frac{\zeta' _{\bar j} }{\sqrt {\xi' _{\bar j} } }z\exp
\left(  - a_{i} z - \xi' _{\bar j} b_{i'} z \right)K_1 \left(
2z\sqrt {\xi' _{\bar j} a_{i} b_{i'} } \right),
\end{align}
\begin{align}
f_{21}  = \frac{{\zeta _j }}{{\sqrt {\xi _j } }}z\exp \left( { -
a_{i} \xi _j z - b_{i'} z} \right)K_1 \left( {2z\sqrt {\xi _j a_{1i}
b_{i'} } } \right),
\end{align}
and
\begin{align}
f_{22}  = \frac{{\zeta' _{\bar j}\zeta _j  }}{{\sqrt {\xi' _{\bar
j}\xi _j } }}z\exp \left( { - \xi _j a_{i} z - \xi' _{\bar j} b_{i'}
z} \right)K_1 \left( {2z\sqrt {\xi _j \xi' _{\bar j}a_{i}  b_{i'} }
} \right).
\end{align}

In addition, $\{j,\overline j\}=\{1,2\}~\mbox{or}~\{2,1\}$,
$\sum\limits_{A_{t'} }$ is the abbreviation of
$\sum\limits_{\scriptstyle A_{t'} \subseteq \left\{ {1, \ldots ,N}
\right\}\backslash i' \hfill \atop
  \scriptstyle \left| {A_{t'} } \right| = t' \hfill}$, $\xi'_j$ and $\zeta'_j$ can be
obtained by $\xi_j$ and $\zeta_j$, respectively, by substituting $i,
l, A_t$ in \eqref{Eq:xi} and \eqref{Eq:zeta} with $i', l', A_{t'}$,
respectively, and $K_1\left(\cdot\right)$ is the first order
modified Bessel function of the second kind\cite{Abramowitz}.

\emph{\textbf{Proof: }}The derivation is given in Appendix D.$\hfill
\blacksquare$

\subsection{Analytical Average SER Analysis of the RS Scheme \eqref{Eq:outdatedRS} with Outdated CSI}

For many common linear modulation formats, the average SER can be
obtained by \cite{Soysa2012}
\begin{align}\label{Eq:SER}
\overline {SER}  = \alpha \mathbb{E}  \left[Q\left(\sqrt {\beta
\gamma } \right)\right]= \frac{\alpha }{{\sqrt {2\pi }
}}\int\limits_0^\infty {F_\gamma \left(\frac{{t^2 }}{\beta
}\right)e^{ - \frac{{t^2 }}{2}} } dt
\end{align}
where $\gamma$ is the instantaneous received SNR,
$Q\left(\cdot\right)$ is the Gaussian
\emph{Q}-Function\cite{Abramowitz}, and $\left(\alpha,\beta\right)$
are decided by the modulation formats\cite{Soysa2012}, e.g.,
$\left(\alpha,\beta\right)=\left(1,2\right)$ for BPSK.

\emph{\textbf{Proposition 2:}} Substituting Proposition $1$ into
(\ref{Eq:SER}), the average SER expression of $S_j$ is obtained
\begin{align}\label{Eq:SER1}
&\overline {SER} _j  = \frac{\alpha }{2} - \frac{{3\sqrt 2 \pi \alpha \sqrt \beta  }}{2}\sum\limits_{i = 1}^N {\sum\limits_{t = 0}^{N - 1} {\sum\limits_{A_t } {\sum\limits_{i' = 1}^N {\sum\limits_{t' = 0}^{N - 1} {\sum\limits_{A_{t'} } {\left( { - 1} \right)^{t + t'} \left( {1 \hspace{-0.8mm} +\hspace{-0.8mm}  \sum\limits_{l \in A_t } {\frac{{\sigma _{\bar ji}^2 }}{{\sigma _{l}^2 }}} } \right)^{ - 1} \left( {1 \hspace{-0.8mm} +\hspace{-0.8mm}  \sum\limits_{l' \in A_{t'} } {\frac{{\sigma _{ji'}^2 }}{{\sigma _{l'}^2 }}} } \right)^{ - 1} a_{i} b_{i'} } } } } } }  \notag\\
&\hspace{84mm}\times \left( {g_{11}  + g_{12}  + g_{21}  + g_{22} }
\right)
\end{align}
where
\begin{align}
g_{11}  = \left[ {\left( {\sqrt {a_{i} }  + \sqrt {b_{i'} } }
\right)^2  + \frac{\beta }{2}} \right]^{ - \frac{5}{2}} F\left(
{\frac{5}{2},\frac{3}{2};2;\frac{{\left( {\sqrt {a_{i} }  - \sqrt {
b_{i'} } } \right)^2  + \frac{\beta }{2}}}{{\left( {\sqrt { a_{i} }
+ \sqrt {b_{i'} } } \right)^2  + \frac{\beta }{2}}}} \right),
\end{align}
\begin{align}
g_{12}  = \zeta' _{\bar j} \left[ {\left( {\sqrt {a_{i} } + \sqrt
{\xi' _{\bar j} b_{i'} } } \right)^2 + \frac{\beta }{2}} \right]^{ -
\frac{5}{2}} F\left( {\frac{5}{2},\frac{3}{2};2;\frac{{\left( {\sqrt
{a_{i} } - \sqrt {\xi' _{\bar j} b_{i'} } } \right)^2  + \frac{\beta
}{2}}}{{\left( {\sqrt { a_{i} }  + \sqrt {\xi' _{\bar j} b_{i'} } }
\right)^2  + \frac{\beta }{2}}}} \right),
\end{align}
\begin{align}
g_{21}  = \zeta _j \left[ {\left( {\sqrt {\xi _j  a_{i} }
 + \sqrt { b_{i'} } } \right)^2  +
\frac{\beta }{2}} \right]^{ - \frac{5}{2}} F\left(
{\frac{5}{2},\frac{3}{2};2;\frac{{\left( {\sqrt {\xi _j a_{i} } -
\sqrt { b_{i'} } } \right)^2  + \frac{\beta }{2}}}{{\left( {\sqrt
{\xi _j a_{i} }  + \sqrt { b_{i'} } } \right)^2  + \frac{\beta
}{2}}}} \right),
\end{align}
and
\begin{align}
g_{22}  = \zeta _j \zeta' _{\bar j} \left[ {\left( {\sqrt {\xi _j
a_{i} }  + \sqrt {\xi' _{\bar j} b_{i'} } } \right)^2 + \frac{\beta
}{2}} \right]^{ - \frac{5}{2}} F\left(
{\frac{5}{2},\frac{3}{2};2;\frac{{\left( {\sqrt {\xi _j a_{i} }  -
\sqrt {\xi' _{\bar j} b_{i'} } } \right)^2 + \frac{\beta
}{2}}}{{\left( {\sqrt {\xi _j a_{i} }  + \sqrt {\xi' _{\bar j}
 b_{i'} } } \right)^2  + \frac{\beta }{2}}}} \right).
\end{align}

In addition, $F\left(a,b,;c;z\right)$ is the Confluent
Hypergeometric function\cite{Abramowitz}.

\emph{\textbf{Proof: }}The derivation is given in Appendix D.$\hfill
\blacksquare$

Proposition~2 provides the generalized average SER expression.
However, this expression of average SER in Proposition 2 is too
complicated, thus we resort to the asymptotic analysis to simplify
the expression.

\subsection{Asymptotic Average SER Analysis in high SNR}

 \emph{\textbf{Corollary 1:}} By Lemma 3 and Proposition 2, the asymptotic average SER of $S_j$, $\left\{j,\overline j\right\}=\left\{1,2\right\}~\mbox{or}~\left\{2,1\right\}$, in high
SNR is

$\left(1\right)$ If CSI is perfect, i.e., $f_{d_{ji}}=0$
\begin{align}\label{Eq:cor11}
&\overline {SER}_j^{\infty}  = \frac{\alpha }{{2\sqrt \pi  }}\left( {\frac{2}{\beta }} \right)^N \frac{{\Gamma \left( {1/2 + N} \right)}}{{\Gamma \left( {N + 1} \right)}} \sum\limits_{i = 1}^N {\sum\limits_{t = 0}^{N - 1} {\sum\limits_{A_t } {\left( { - 1} \right)^{N + t + 1} } } } \notag\\
&\hspace{30mm}\times \left[ {\left( {\sum\limits_{l \in A_t }
{\frac{{\sigma _{ji}^2 }}{{\sigma _{l}^2 }}} } \right)^{N - 1}\left(
{\frac{1}{{\psi_r \sigma _{ji}^2 }}} \right)^N + \left(
{\sum\limits_{l \in A_t } {\frac{{\sigma _{\bar j i}^2 }}{{\sigma
_{l}^2 }}} } \right)^{N - 1}\left( {\frac{1}{{\psi_h \sigma _{\bar j
i}^2 }}} \right)^N} \right]
\end{align}
where $\Gamma\left(\cdot\right)$ is the Gamma
function\cite{Abramowitz}.

$\left(2\right)$ If CSI is outdated, i.e., $f_{d_{ji}}>0$
\begin{align}\label{Eq:cor12}
\overline {SER_j } ^\infty   = \frac{\alpha }{{2\beta
}}\sum\limits_{i = 1}^N {\sum\limits_{t = 0}^{N - 1}
{\sum\limits_{A_t } {\left( { - 1} \right)^t } } } \left[\frac{{1 +
\zeta _j }}{{1 + \sum\limits_{l \in A_t } {\frac{{\sigma _{\bar
ji}^2 }}{{\sigma _l^2 }}} }}\frac{1}{{\psi _r \sigma _{ji}^2 }}{\rm{
}} + \frac{{1 + \zeta _{\bar j} }}{{1 + \sum\limits_{l \in A_t }
{\frac{{\sigma _{ji}^2 }}{{\sigma _l^2 }}} }}\frac{1}{{\psi _h
\sigma _{\bar ji}^2 }}\right].
\end{align}

\emph{\textbf{Proof: }}The derivation is given in Appendix E.$\hfill
\blacksquare$

Furthermore, it is worthy of noting that the previous analytical
expressions, i.e., from Lemma~1 to Corollary~1, are all obtained
under the generalized network structure, i.e., the variances
$\sigma_{ji}^2$ and the correlation coefficients $\rho_{ji}$ for
different channels are different. If the network structure is
symmetric, i.e., $\sigma_{ji }^2=1$ and $\rho_{ji}=\rho$, the
previous expressions can be further simplified by
$\sum\limits_{i=1}^N \to N$, $\sum\limits_{A_t} \to \binom{N-1}{t}$,
and $\sum\limits_{l \in A_t} \to t$.

\subsection{Diversity Analysis of Single and Multiple RS Schemes with Outdated CSI}

With the aid of the asymptotic SER expressions, diversity order $d$,
which implies the slope of SER in log-log scale when SNR approaches
infinity\cite{Zheng2003}, satisfies

\emph{\textbf{Corollary 2:}}
\begin{equation}
d=
\begin{cases}
  N, ~\text{CSI is perfect} ; \\
  1, ~\text{CSI is outdated}.
\end{cases}
\end{equation}

 \emph{\textbf{Proof: }} The proof is given in
Appendix E.$\hfill \blacksquare$

Corollary~2 reveals that the diversity order of the single relay
selection scheme \eqref{Eq:outdatedRS} degrades to one from full
diversity, once the CSI is outdated.

To compensate the diversity loss, a multiple RS scheme with outdated
CSI by selecting the best $K$ relays from the $N$ available relays
is proposed.

Assuming $\varphi^{(K)}$ to be the $K$th largest value among the set
$\big\{\varphi_i\buildrel \Delta \over =\min \big(|\hat
h_{1i}|^2,|\hat h_{2i}|^2\big), i=1,\ldots,N\big\}$, the relay
$R_i$, $i=1,\ldots,N$, is selected if and only if
\begin{align}\label{Eq:MR}
\varphi_i \ge \varphi^{(K)}.
\end{align}

The selected $K$ relays can forward the signals in the orthogonal
resources, and the maximal ratio combing is adopted at the sources.

\emph{\textbf{Proposition 3:}} Selecting the best $K$ relays from
$N$ available relays
 by the RS scheme \eqref{Eq:MR} and using the maximal-ratio combining, the diversity order
 is
\begin{equation}
d=
\begin{cases}
  N, ~\text{CSI is perfect} ; \\
  K, ~\text{CSI is outdated}.
\end{cases}
\end{equation}

 \emph{\textbf{Proof: }} The proof is given in
Appendix F.$\hfill \blacksquare$

Therefore, increasing the number of selected relay $K$ results in
the improvement of diversity order in high SNR, thus the average SER
performance also gets improved.

It is worthy point out that from the perspective of diversity order,
multiple RS is not better than single RS when CSI is perfect,
because they both can achieve the full diversity, and single RS only
exploits one relay \cite{Jing2009TWC}. Nevertheless, multiple RS can
improve the diversity order with outdated CSI, in comparison with
single RS \cite{Dong2012}.

\section{Simulation Results and Discussion}

In this section, Monte-Carlo simulations are provided to validate
the preceding analysis and to highlight the performance of
bidirectional AF RS with outdated CSI. Without loss of generality,
the average SER of the simulation results only concern about $S_1$
under BPSK modulation. Moreover, the variance of the channel
satisfies $\sigma_{{ji}}^2=1$, and the Doppler spread of the channel
satisfies $f_{d_{ji}}=f_d$, $i=1,\ldots,N$, and $j=1,2$.

Figs.~2-5 investigate the performance of single RS scheme
\eqref{Eq:outdatedRS}, in which each source and each relay are
assumed to have the same transmit powers, i.e., $p_s=p_r=P_0$.

\begin{figure}[h!]
\centering
\includegraphics[height=3.6in,width=4.0in]{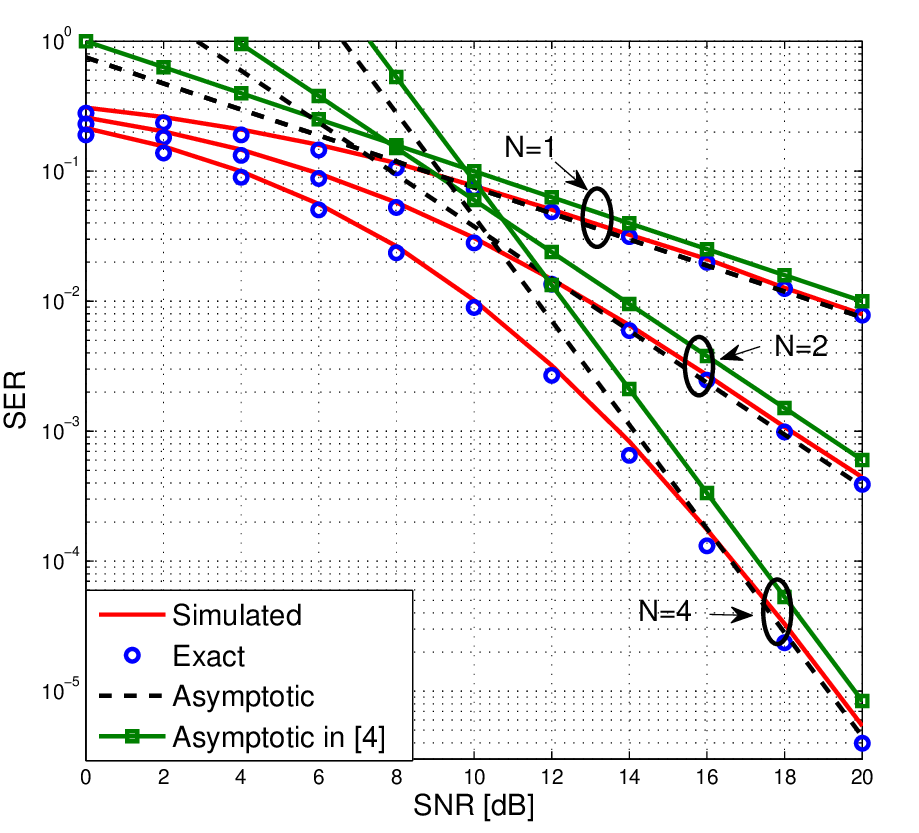}
\caption{SER of $S_1$ when CSI is perfect and $N=1,2,4$.}
\label{fig:perfect}
\end{figure}

In Fig.~\ref{fig:perfect}, the simulation and the analytical SER of
bidirectional RS are provided with perfect CSI, i.e, $f_dT_d=0$,
when the number of relays $N=1,2,4$. The x-axis of this figure is
$\mbox{SNR}=P_0/\sigma_n^2$ in dB. This figure reveals that
increasing the number of relays can reduce the average SER, because
the diversity order is $N$ when CSI is perfect, which satisfies the
result of Corollary 2. From this figure, the exact SER expression of
Proposition 2 is verified when CSI is perfect, in which the exact
analytical expression of SER tightly matches with the simulation
results than the previous researches\cite{Song2011}, and the
asymptotic results obtained from Corollary 1 also converges to the
simulation results in high SNR.

\begin{figure}[h!]
\centering
\includegraphics[height=3.6in,width=4.0in]{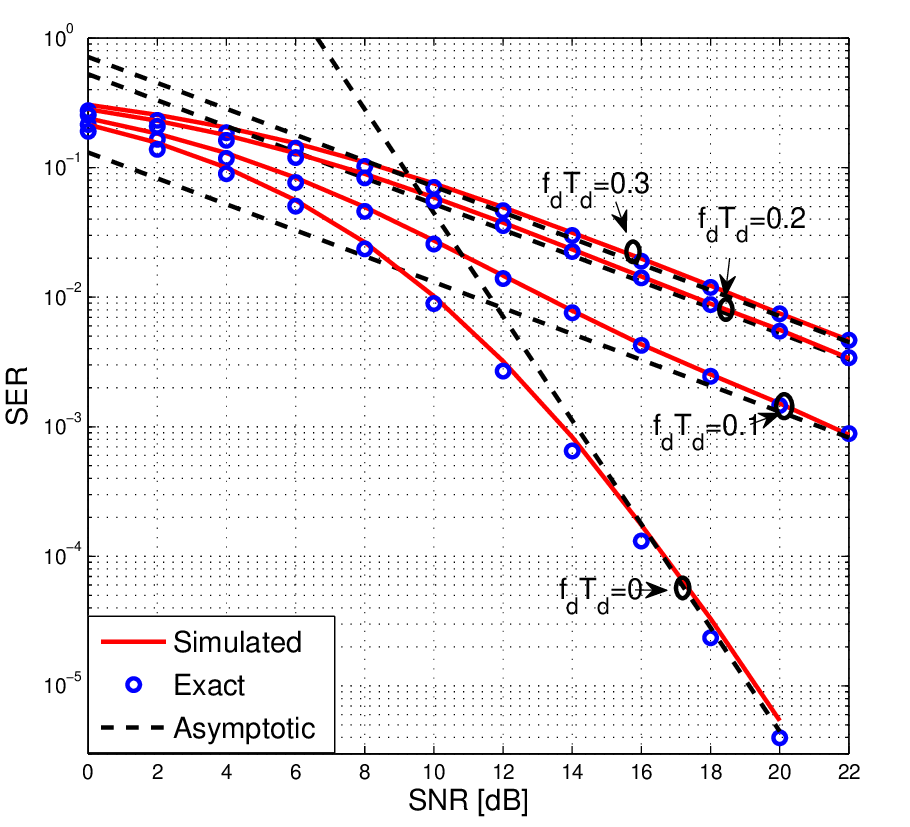}
\caption{SER of $S_1$ when CSI is outdated and $N=4$.}
\label{fig:pf}
\end{figure}

Fig.~\ref{fig:pf} studies the impact of outdated CSI on the SER
performance when $N=4$ and $f_dT_d=0,0.1,0.2,0.3$. The x-axis of
this figure is $\mbox{SNR}=P_0/\sigma_n^2$ in dB. Different lines
are provided under different $f_dT_d$, where larger $f_dT_d$ means
CSI is severely outdated, whereas smaller $f_dT_d$ means CSI is
slightly outdated, and especially $f_dT_d=0$ means CSI is perfect.
The figure verifies the expressions of Proposition 2 and Corollary 1
when CSI is outdated. The figure also presents the adverse effect of
outdated CSI on the performance: if and only if CSI is perfect, the
diversity order is $N$; however, once CSI is outdated, the
performance degrades greatly that the diversity order reduces to
$1$, which satisfies Corollary $2$. The qualitative explanation of
the phenomenon is that once CSI is outdated, it is quite possible
that the worst relay can be selected, and hence diversity order is
$1$. Furthermore, although diversity order is the same for any
$f_dT_d>0$, the performance loss is smaller for smaller $f_dT_d$.
Specifically, the gap of SER between $f_dT_d=0.1$ and $f_dT_d=0.2$
is about $4$ dB in high SNR.

\begin{figure}[h!]
\centering
\includegraphics[height=3.6in,width=4.0in]{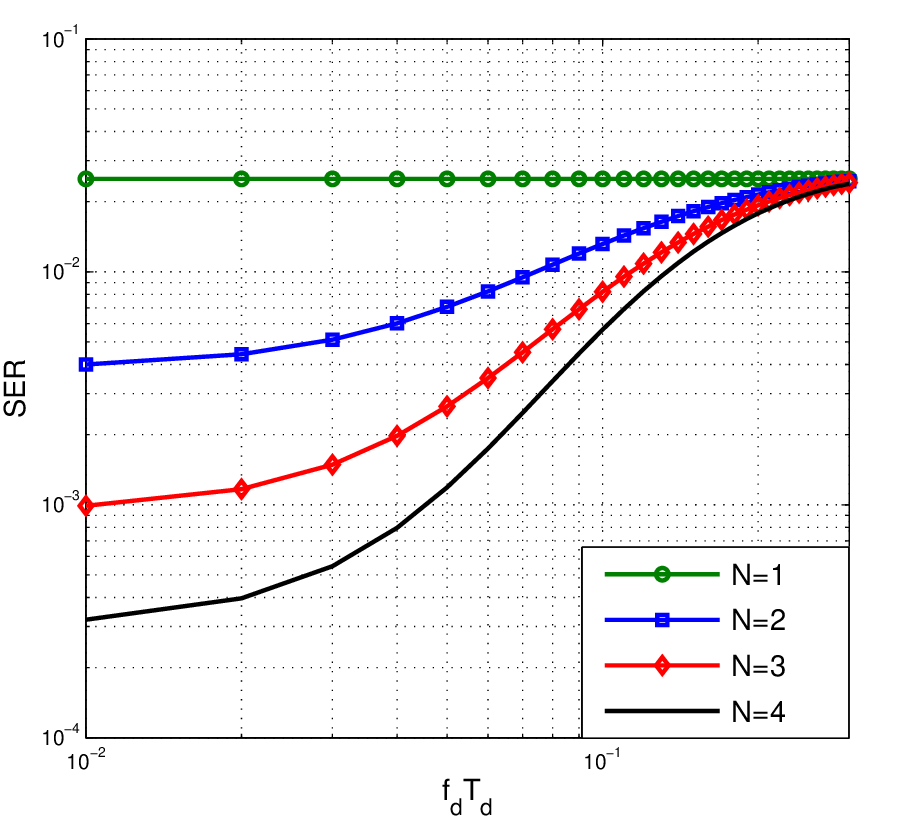}
\caption{The impact of $f_dT_d$ on the SER when $\mbox{SNR}=15$~dB
and $N=1,2,3,4$.} \label{fig:fd}
\end{figure}

Fig.~\ref{fig:fd} investigates the impact of  $f_dT_d$ on SER when
$\mbox{SNR}=15$~dB and the number of relays $N=1,2,3,4$. As the
figure reveals, the SER gets worse as the CSI becomes severely
outdated. During the range of small $f_dT_d$, the SER of larger $N$
still have significant gain over the SER of smaller $N$, whereas all
the curves approach to the performance of $N=1$ as $f_dT_d$
increases. This indicates that with severely outdated CSI, no
significant performance gain can be achieved by deploying more
relays. Therefore, the RS scheme \eqref{Eq:outdatedRS} behaves as
the random RS, when the CSI is severely outdated.

\begin{figure}[h!]
\centering
\includegraphics[height=3.6in,width=4.0in]{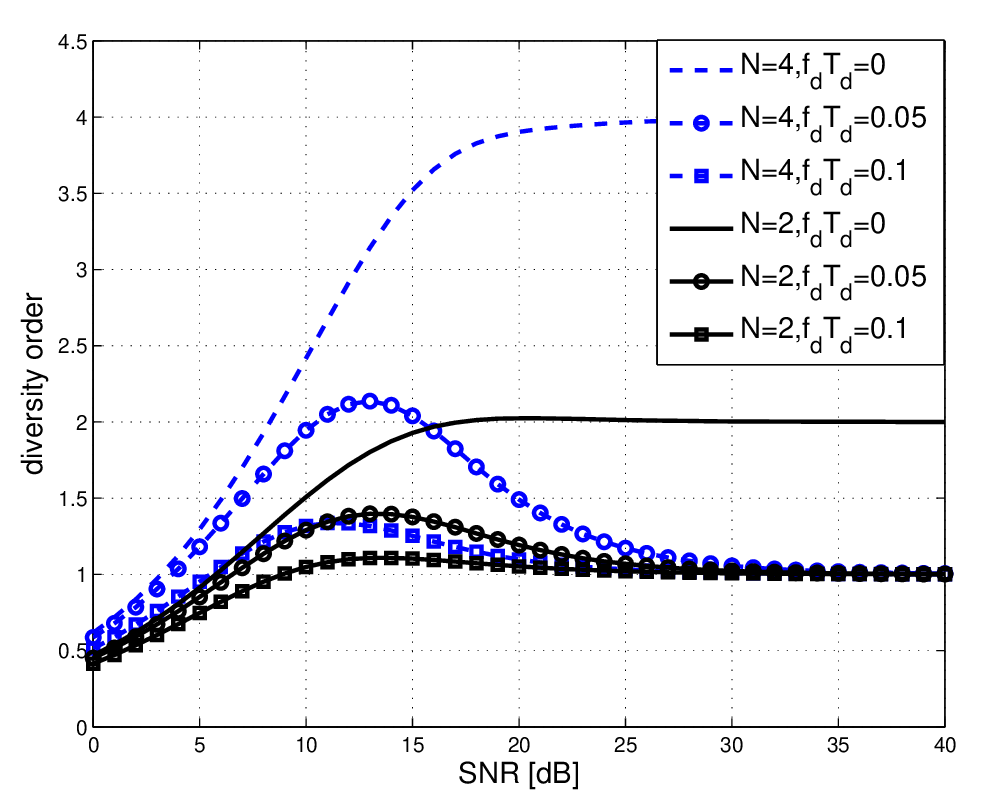}
\caption{Diversity order versus finite SNR under different $f_dT_d$
and different $N$.} \label{fig:do}
\end{figure}

Fig.~\ref{fig:do} plots the diversity order of finite SNR
\cite{Seyi2011} when $f_dT_d=0,0.05,0.1$ and $N=2,4$. This figure
verifies the Corollary~2, i.e., once CSI is outdated, the diversity
order when SNR approaches infinity degrades to one from full
diversity, regardless of $N$ and $f_dT_d$. However, the diversity
order of finite SNR is different for different $f_dT_d$ and
different $N$. Specifically, Fig.~\ref{fig:do} reveals that the
outdated CSI has little impact on the diversity order of low SNR,
which is also verified by the Fig.~\ref{fig:pf}, in which the SER
curves with outdated CSI maintain their slopes for low SNR.
Nevertheless, due to the great impact of outdated CSI on high SNR,
the diversity order converges to one as the SNR grows infinitely
large. This phenomenon illustrates that, although it is impossible
to achieve full diversity when SNR approaches infinity, the
diversity order is preserved for an SNR interval which increases as
$f_dT_d$ decreases. Furthermore, this figure also reveals that the
diversity order with larger $N$ is no less than the diversity order
with smaller $N$ all over the SNR. For instance, when $f_dT_d=0.05$,
the diversity order of $\mbox{SNR}=10$ dB with $N=2$ is about $1.3$,
whereas it increases to $2.1$, when $N=4$.

\begin{figure}[h!]
\centering
\includegraphics[height=3.6in,width=4.0in]{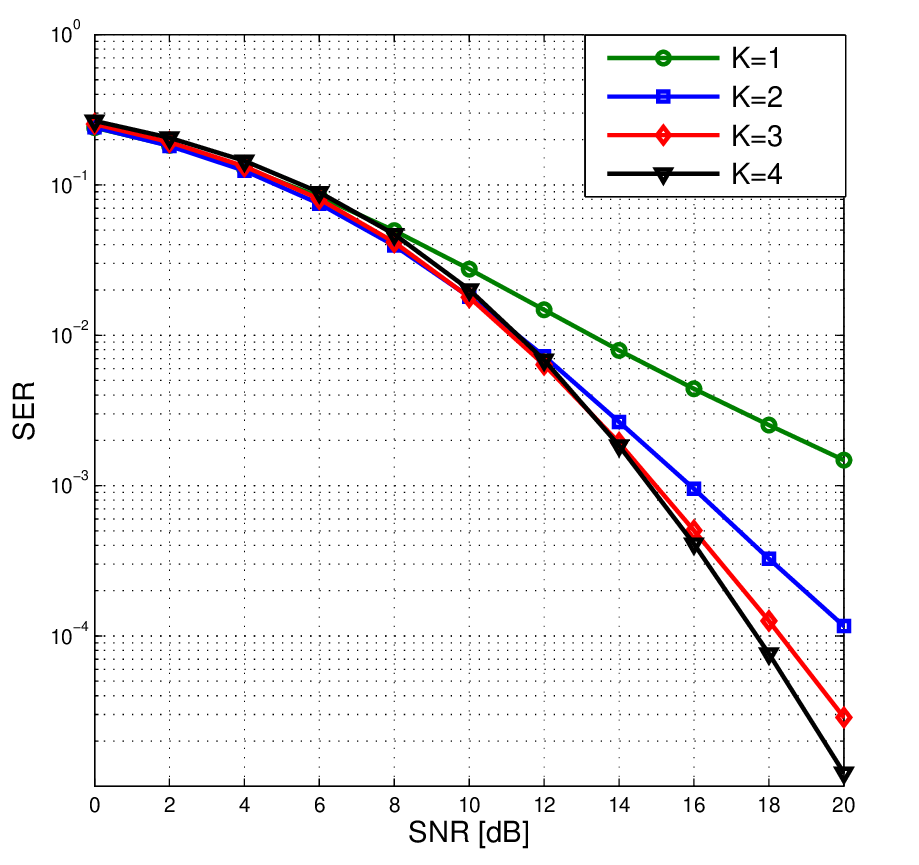}
\caption{Multiple relay selection under outdated CSI when $N=4$,
$K=1,2,3,4$, and $f_dT_d=0.1$.} \label{fig:MRS}
\end{figure}

Fig.~\ref{fig:MRS} studies the performance of multiple RS when $N=4$
and $f_dT_d=0.1$. The best $K$ relays, $K=1,2,3,4$, are selected
according to \eqref{Eq:MR} and the maximal ratio combining is
adopted. For the sake of fairness, the total power of the selected
relays is assumed to be the same, regardless of $K$. In addition,
the total power is allocated equally among the selected relays, i.e,
$p_r=P_0/K$. We also assume that $p_s=P_0$, and the x-axis of this
figure is $\mbox{SNR}=P_0/\sigma_n^2$ in dB. From this figure, we
find that in low SNR, the SER performance for different $K$ is
almost the same, because the performance in low SNR is
power-limited, and the total power is the same for different $K$.
However, in high SNR, increasing the number of selected relays $K$
will significantly improve the SER performance, because the
diversity order is $K$ with outdated CSI, which satisfies the
analysis of Proposition~3. Therefore, multiple RS is a feasible
solution to improve the diversity loss caused by the outdated CSI.

\section{Conclusions}
The effect of outdated CSI on the SER performance of bidirectional
AF RS has been investigated in this paper. For the single RS, the
distribution of end-to-end SNR, the analytical and asymptotic
expressions of SER are derived in a closed form, and verified by
simulations. The effect of the number of relays and the correlation
coefficient of outdated CSI are investigated. The results reveal
that the SER performance of bidirectional RS is highly dependent on
the outdated CSI. Specifically, the diversity order reduces to one
from full diversity, once CSI is outdated. Furthermore, a multiple
RS scheme is proposed, and the diversity order with outdated CSI is
analyzed, which proves that the multiple RS scheme can compensate
the diversity loss caused by outdated CSI.

\section*{Appendix A: Proof of Lemma~1}

According to the inequality\cite{Upadhyay2011}
${xy}/{\left(x+y\right)} \le \min \left(x,y\right), x,y>0$, we have
\begin{align}\label{Eq:A3}
\min \left( {\gamma _{1i} ,\gamma _{2i} } \right) \le \min \left[
{\min \left( {\psi _r \left| {h_{1i} } \right|^2 ,\psi _h \left|
{h_{2i} } \right|^2 } \right),\min \left( {\psi _h \left| {h_{1i} }
\right|^2 ,\psi _r \left| {h_{2i} } \right|^2 } \right)} \right].
\end{align}

Furthermore, by the fact $\psi_r>\psi_h\buildrel \Delta
\over={\psi_r\psi_s}/\left({\psi_r+\psi_s}\right)$ and discussion
under different conditions, we will verify that
\begin{align}\label{Eq:AAA}
\min \left[ {\min \left( {\psi _r \left| {h_{1i} } \right|^2 ,\psi
_h \left| {h_{2i} } \right|^2 } \right),\min \left( {\psi _h \left|
{h_{1i} } \right|^2 ,\psi _r \left| {h_{2i} } \right|^2 } \right)}
\right] = \psi _h \min \left( {\left| {h_{1i} } \right|^2 ,\left|
{h_{2i} } \right|^2 } \right).
\end{align}

The process of verifying \eqref{Eq:AAA} is listed as follows, where
$\psi_h<\psi_r$:

 (i) If
$\psi_r|h_{1i}|^2<\psi_h|h_{2i}|^2$, we have
$|h_{1i}|^2<|h_{2i}|^2$, thus $\psi_h|h_{1i}|^2<\psi_r|h_{2i}|^2$.
Therefore, we have
\begin{align}\label{Eq:i}
&I \buildrel \Delta \over = \min \left[ {\min \left( {\psi _r |h_{1i} |^2 ,\psi _h |h_{2i} |^2 } \right),\min \left( {\psi _h |h_{1i} |^2 ,\psi _r |h_{2i} |^2 } \right)} \right]\notag \\
&= \min \left( {\psi _r |h_{1i} |^2 ,\psi _h |h_{1i} |^2 } \right) =
\psi _h |h_{1i} |^2  = \psi _h \min \left( {|h_{1i} |^2 ,|h_{2i} |^2
} \right)
\end{align}

(ii) If $\psi_r|h_{1i}|^2 \ge \psi_h|h_{2i}|^2$,  $ \min \{ \psi _r
|h_{1i} |^2 ,\psi _h |h_{2i} |^2\}= \psi _h |h_{2i} |^2 $.

(iia) Under situation (ii) and if
$\psi_h|h_{1i}|^2<\psi_r|h_{2i}|^2$, we have $ \min \{ \psi _h
|h_{1i} |^2 ,\psi _r |h_{2i} |^2 \}=\psi _h |h_{1i} |^2 $, thus
\begin{align}\label{Eq:ia}
I = \psi _h \min \left( {|h_{1i} |^2 ,|h_{2i} |^2 } \right).
\end{align}

(iib) Under situation (ii) and if $\psi_h|h_{1i}|^2 \ge
\psi_r|h_{2i}|^2$, we have $ \min \{ \psi _h |h_{1i} |^2 ,\psi _r
|h_{2i} |^2 \}=\psi _r |h_{2i} |^2 $ and $|h_{1i}|^2 > |h_{2i}|^2$,
thus
\begin{align}\label{Eq:ib}
I = \min \left( {\psi _h |h_{2i} |^2 ,\psi _r |h_{2i} |^2 } \right)
= \psi _h |h_{2i} |^2  =\psi _h  \min \left( {|h_{1i} |^2 , |h_{2i}
|^2 } \right).
\end{align}

In summary of \eqref{Eq:i}, \eqref{Eq:ia}, and \eqref{Eq:ib},
\eqref{Eq:AAA} is achieved. Therefore, Lemma~1 is verified according
to \eqref{Eq:A3} and \eqref{Eq:AAA}.

\section*{Appendix B: Proof of Lemma~2 }
Inspired by \cite{Michalopoulos2012}, the PDF of $|h_{1k}|^2$ can be
expanded as
\begin{align}\label{Eq:lemma21}
&f_{\left| {h_{1k} } \right|^2 } \left( z \right) \buildrel (a) \over = \frac{d}{{dz}}\sum\limits_{i = 1}^N {\Pr \left\{ {\left| {h_{1i} } \right|^2  < z \cap k = i} \right\}}  \notag\\
&= \frac{d}{{dz}}\sum\limits_{i = 1}^N {\int\limits_0^z {\int\limits_0^\infty  {f_{\left| {h_{1i} } \right|^2 ,\left| {\hat h_{1i} } \right|^2 } \left( {x,y} \right)} } \Pr \left\{ {k = i|\left| {h_{1i} } \right|^2  = x,\left| {\hat h_{1i} } \right|^2  = y} \right\}dxdy}  \notag\\
& = \sum\limits_{i = 1}^N {\int\limits_0^\infty  {f_{\left| {h_{1i} } \right|^2 |\left| {\hat h_{1i} } \right|^2 } \left( {z|y} \right)f_{\left| {\hat h_{1i} } \right|^2 } \left( y \right)\Pr \left\{ {k = i|{\left| {\hat h_{1i} } \right|^2 }  = y} \right\}dy} }  \notag\\
&\buildrel (b)\over = \sum\limits_{i = 1}^N {\int\limits_0^\infty  {f_{\left| {h_{1i} } \right|^2 |\left| {\hat h_{1i} } \right|^2 } \left( {z|y} \right)f_{\left| {\hat h_{1i} } \right|^2 } \left( y \right)\Pr \left\{ {\left| {\hat h_{1i} } \right|^2 \hspace{-0.8mm}  \le\hspace{-0.8mm}  \left| {\hat h_{2i} } \right|^2 |\left| {\hat h_{1i} } \right|^2 \hspace{-0.8mm}  = \hspace{-0.8mm} y} \right\}\Pr \left\{ {k = i|\left| {\hat h_{1i} } \right|^2 \hspace{-0.8mm}  =\hspace{-0.8mm}  y,\left| {\hat h_{1i} } \right|^2 \hspace{-0.8mm}  \le \hspace{-0.8mm} \left| {\hat h_{2i} } \right|^2 } \right\}dy} } \notag \\
&+ \sum\limits_{i = 1}^N {\int\limits_0^\infty  {f_{\left| {\hat h_{1i} } \right|^2 |\left| {\hat h_{2i} } \right|^2 } \left( {z|y} \right)f_{\left| {\hat h_{1i} } \right|^2 } \left( y \right)\Pr \left\{ {\left| {\hat h_{1i} } \right|^2  \hspace{-0.8mm} >\hspace{-0.8mm}  \left| {\hat h_{2i} } \right|^2 |\left| {\hat h_{1i} } \right|^2 \hspace{-0.8mm}  =\hspace{-0.8mm}  y} \right\}\Pr \left\{ {k = i|\left| {\hat h_{1i} } \right|^2 \hspace{-0.8mm}  =\hspace{-0.8mm}  y,\left| {\hat h_{1i} } \right|^2 \hspace{-0.8mm}  >\hspace{-0.8mm}  \left| {\hat h_{2i} } \right|^2 } \right\}dy} }  \notag\\
&\buildrel (c) \over = \sum\limits_{i = 1}^N {\int\limits_0^\infty  {f_{\left| {h_{1i} } \right|^2 \left| {\hat h_{1i} } \right|^2 } \left( {z|y} \right)f_{\left| {\hat h_{1i} } \right|^2 } \left( y \right)\Pr \left\{ {\left| {\hat h_{1i} } \right|^2  \le \left| {\hat h_{2i} } \right|^2 |\left| {\hat h_{1i} } \right|^2  = y} \right\}I_i \left( y \right)dy} } \notag \\
&+ \sum\limits_{i = 1}^N {\int\limits_0^\infty  {f_{\left| {h_{1i} }
\right|^2 |\left| {\hat h_{1i} } \right|^2 } \left( {z|y}
\right)f_{\left| {\hat h_{1i} } \right|^2 } \left( y \right)\Pr
\left\{ {\left| {\hat h_{1i} } \right|^2  > \left| {\hat h_{2i} }
\right|^2 |\left| {\hat h_{1i} } \right|^2  = y} \right\}I_i \left(
{\left| {\hat h_{2i} } \right|^2 } \right)dy} }
\end{align}
where (a) is satisfied by total probability theorem\cite{Paoulis},
in which the universal set $|h_{1k}|^2<z$ is divided into $N$
disjoint sets $|h_{1i}|^2<z$, $i=1,\ldots,N$; (b) is fulfilled by
the division of two disjoint events, i.e., $ |\hat h_{1i}|^2  >
|\hat h_{2i}|^2$ and $ |\hat h_{1i}|^2 \le |\hat h_{2i}|^2$, and (c)
is decided by the RS scheme \eqref{Eq:outdatedRS}, in which $I_i
\left( x \right) \buildrel \Delta \over = \prod\limits_{q = 1,q \ne
i}^N {\Pr \left\{ {\min \left(   |\hat h_{1i}|^2,  |\hat h_{2i}|^2
\right) \le x} \right\}} $.

According to order statistics\cite{David1970} and\cite[
eq.~(26)]{Hai2011} , $I_i\left(x\right)$ can be expanded as
\begin{align}
I_i \left( x \right) & = \prod\limits_{q=1,q \ne i}^N {\left\{ {1 - \left[ {1 - \Pr \left( {  |\hat h_{1q}|^2  \le x} \right)} \right]\left [ {1 - \Pr \left( {   |\hat h_{2q}|^2  \le x} \right)} \right]} \right\}}  \notag\\
&= \prod\limits_{q=1,q \ne i}^N {\left\{ {1 - \exp \left( { -
\frac{x}{{\sigma _{q}^2 }}} \right)} \right\}} \notag\\
&=1 + \sum\limits_{t = 1}^{N - 1} {\sum\limits_{\scriptstyle A_t
\subseteq \left\{ {1, \ldots ,N} \right\}\backslash i \hfill \atop
  \scriptstyle \left| {A_t } \right| = t \hfill} {\left( { - 1} \right)^t \exp \left( { - x\sum\limits_{l \in A_t } {\frac{1}{{\sigma _{l}^2 }}} } \right)}
}
\end{align}
where $ \frac{1}{{\sigma _{l}^2 }} \buildrel \Delta \over =
\frac{1}{{\sigma _{1l}^2 }} + \frac{1}{{\sigma _{2l}^2 }}$,
$l=1,\ldots,N$.

According to the exponential distribution of $|\hat h_{1i}|^2$ and $
|\hat h_{2i}|^2$, and the conditional PDF
\cite[eq.~(31)]{Michalopoulos2012}
\begin{align}\label{Eq:ccdf}
f_{|h_{ji}|^2 ||\hat h_{ji}|^2 } \left( {z|y} \right) =
\frac{1}{{\left( {1 - \rho _{ji}^2 } \right)\sigma _{ji}^2 }}\exp
\left( { - \frac{{\rho _{ji}^2 y +  z}}{{ \left( {1 - \rho _{ji}^2 }
\right)\sigma _{ji}^2 }}} \right) I_0 \left( {\frac{{2\sqrt {\rho
_{ji}^2 yz} }}{{\left( {1 - \rho _{ji}^2 } \right)\sigma _{ji}^2
 }}} \right),
\end{align}
the PDF of $ | h_{1k}|^2$ is verified by  \cite[eq.
(6.614.3)]{Gradshteyn94}
\begin{align}
\int_0^\infty {\exp \left( - \alpha x\right)I_0 \left(\beta \sqrt x
\right)} dx = \left({1}/{\alpha }\right)\exp \left({{\beta ^2
}}/\left({{4\alpha }}\right)\right),
\end{align}
and the PDF of $| h_{2k}|^2$ can be verified similarly.

\section*{Appendix C: Proof of Lemma~3 }
According to the fact \cite[ eq.~(26)]{Hai2011}, i.e.,
\begin{align}\label{Eq:lemma33}
\prod\limits_{q = 1,q \ne i}^N {\left\{ {1 - \exp \left( { -
\frac{x}{{\sigma _{q}^2 }}} \right)} \right\}} {\rm{ =
}}\sum\limits_{t = 0}^{N - 1} {\sum\limits_{A_t } {\left( { - 1}
\right)^t \exp \left( { - x\sum\limits_{l \in A_t }
{\frac{1}{{\sigma _{l}^2 }}} } \right)} },
\end{align}
we have
\begin{align}\label{Eq:lemma34}
\sum\limits_{i = 1}^N {\frac{1}{{\sigma _{i}^2 }}} \exp \left( { -
\frac{x}{{\sigma _{i}^2 }}} \right)\prod\limits_{q = 1,q \ne i}^N
{\left\{ {1 - \exp \left( { - \frac{x}{{\sigma _{q}^2 }}} \right)}
\right\}} =\sum\limits_{i = 1}^N {\frac{1}{{\sigma _{i}^2 }}}
\sum\limits_{t = 0}^{N - 1} {\sum\limits_{A_t } {\left( { - 1}
\right)^t \exp \left( { - \sum\limits_{l \in A_t } {\frac{x}{{\sigma
_{l}^2 }} - \frac{x}{{\sigma _{i}^2 }}} } \right)} }.
\end{align}
By integrating both sides of \eqref{Eq:lemma34} from $0$ to
$\infty$, the formula can be expressed as
\begin{align}\label{Eq:lemma35}
\prod\limits_{i = 1}^N {\left\{ {1 - \exp \left( { -
\frac{x}{{\sigma _{i}^2 }}} \right)} \right\}} \bigg|_0^\infty =
\sum\limits_{i = 1}^N {\frac{1}{{\sigma _{i}^2 }}} \sum\limits_{t =
0}^{N - 1} {\sum\limits_{A_t } {\left( { - 1} \right)^t \frac{{\exp
\left( { - \sum\limits_{l \in A_t } {\frac{x}{{\sigma _{l}^2 }} -
\frac{x}{{\sigma _{i}^2 }}} } \right)}}{{ - \sum\limits_{l \in A_t }
{\frac{1}{{\sigma _{l}^2 }} - \frac{1}{{\sigma _{i}^2 }}} }}} }
\bigg|_0^\infty
\end{align}
where $f\left( x \right)\big|_0^\infty  = f\left( \infty  \right) -
f\left( 0 \right)$, thus the first equation in Lemma 3 is proved.

Substituting $x$ with $0$ into \eqref{Eq:lemma33}, the second
equation in Lemma~3 is proved when $k=0$.

Differentiating \eqref{Eq:lemma33}, we have
\begin{align}\label{Eq:lemma36}
\sum\limits_{k = 1,k \ne i}^{k = N} {\frac{1}{{\sigma _{k}^2 }}}
\prod\limits_{q = 1,q \ne k,i}^N {\left\{ {1 - \exp \left( { -
\frac{x}{{\sigma _{q}^2 }}} \right)} \right\}}  = \sum\limits_{t =
1}^{N - 1} {\sum\limits_{A_t } { \left( {\sum\limits_{l \in A_t }
{\frac{\left( { - 1} \right)^{t+1}}{{\sigma _{l}^2 }}} } \right)\exp
\left( { -\sum\limits_{l\in A_t } {\frac{ x}{{\sigma _{l}^2 }}} }
\right)} },
\end{align}
then substituting $x$ with $0$, the second equation in Lemma 3 is
proved when $k=1$.

It is easily verified that for $k=2,\ldots,N-2$, the second equation
in Lemma 3 is achieved by differentiating \eqref{Eq:lemma36}
continually and then substituting $x$ with $0$ subsequently.

\section*{Appendix D: Proof of Proposition~1 and Proposition~2}
The received SNR of $S_1$ is
$\gamma_{1k}=\Omega_1\Omega_2/\left(\Omega_1+\Omega_2\right)$, where
$\Omega_1=\psi_r  | h_{1k}|^2$ and $\Omega_2=\psi_h | h_{2k}|^2$.
Therefore, the CDF of $\gamma_{1k}$ can be written as
\begin{align}\label{Eq:FF}
F_{\gamma _{1k} } \left( z \right) &= \Pr \left\{ {\frac{{\Omega _1
\Omega _2 }}{{\Omega _1  + \Omega _2 }} < z} \right\} \notag\\
& = \Pr \left\{ {\left( {\Omega _2  - z} \right)\Omega _1  < z\Omega
_2 ,\Omega _2  > z} \right\} + \Pr \left\{ {\left( {\Omega _2  - z}
\right)\Omega _1  < z\Omega _2 ,\Omega _2  \le z} \right\}
\notag\\
&= \int_z^\infty  {F_{\Omega _1 } \left( {\frac{{zx}}{{x - z}}}
\right)} f_{\Omega _2 } \left(x\right)dx + \int_0^z {\left[ {1 -
F_{\Omega _1 } \left( {\frac{{zx}}{{x - z}}} \right)} \right]}
f_{\Omega _2 } \left( x \right)dx \notag\\
 &= 1 - \int\limits_0^\infty  {f_{\Omega _2 } \left( {x + z}
\right)\left[ {1 - F_{\Omega _1 } \left( {z + \frac{{z^2 }}{x}}
\right)} \right]} dx.
\end{align}

Substituting $ \int\limits_0^\infty {\exp \left( - mx - nx^{ - 1}
\right)} dx = 2\sqrt {{n}/{m}} K_1 \left(2\sqrt {mn} \right)$
\cite[eq. (3.324)]{Gradshteyn94} into (\ref{Eq:FF}), Proposition 1
can be proved by Lemma 2.

Applying Proposition 1 and (\ref{Eq:SER}), the exact average SER of
$S_1$ can be obtained by \cite[eq. (6.621.3)]{Gradshteyn94}
\begin{align}\label{}
\int_0^\infty {x^{\mu  - 1} } e^{ - \alpha x} K_\nu  \left(\beta
x\right)dx = \frac{{\sqrt \pi \left(2\beta \right)^\nu
}}{{\left(\alpha \hspace{-0.8mm} + \hspace{-0.8mm}\beta \right)^{\mu
+ \nu } }}\frac{{\Gamma \left(\mu \hspace{-0.8mm} +\hspace{-0.8mm}
\nu \right)\Gamma \left(\mu\hspace{-0.8mm}  -\hspace{-0.8mm} \nu
\right)}}{{\Gamma \left(\mu\hspace{-0.8mm} +\hspace{-0.8mm}
1/2\right)}}F\left(\mu \hspace{-0.8mm}+\hspace{-0.8mm} \nu
,\nu\hspace{-0.8mm} + \frac{1}{2};\mu
\hspace{-0.8mm}+\hspace{-0.8mm} \frac{1}{2};\frac{{\alpha
\hspace{-0.8mm} - \hspace{-0.8mm}\beta }}{{\alpha\hspace{-0.8mm}
+\hspace{-0.8mm} \beta }}\right)
\end{align}
where $\Gamma\left(\cdot\right)$ is the Gamma function, and
$F\left(\cdot\right)$ is the Confluent Hypergeometric
function\cite{Abramowitz}.

The performance of $S_2$ can be verified similarly.

It is noted that the CSI of the selected path is also assumed to be
perfect for data transmission, although the CSI used for relay
selection is outdated. The reasoning behind this assumption lies in
the fact that the time-repetition rates between the above two
processes, i.e., relay selection and data transmission, are
generally different \cite{Michalopoulos2012}. We also assume that
the channel reciprocity is satisfied, i.e., the channel stays
constant during the two phases of communication. This assumption is
reasonable, when the frame length of the two phases is relatively
small, or the correlation coefficient of outdated CSI is relatively
large. Furthermore, this assumption is also made in the previous
research of two-way relay with outdated CSI, such as the user
selection \cite{Fan2012} and the antenna selection
\cite{Amarasuriya2012}. Therefore, we follow this assumption in this
paper.

\section*{Appendix E: Proof of Corollary~1 and Corollary~2}
In high SNR, i.e., $\psi_s,\psi_r\to\infty$, we have $ a_i\to0$ and
$b_i'\to0$. According to $xy/\big(x+y\big)\approx\min \big(x,y\big)$
and the order statistics\cite{David1970}, the CDF of $\gamma_{1k}$
is expressed as
\begin{align}\label{Eq:E1}
F_{\gamma _{1k} } \left( z \right) = 1 - \left[ {1 - F_{\left|
{h_{1k} } \right|^2 } \left( {\frac{z}{{\psi _r }}} \right)}
\right]\left[ {1 - F_{\left| {h_{2k} } \right|^2 } \left(
{\frac{z}{{\psi _h }}} \right)} \right]
\end{align}
where the CDF of $|h_{jk}|^2$ can be obtained by integrating
\eqref{Eq:lemma21}.

If CSI is perfect, i.e., $\rho_{ji}=1$, the CDF of
$\psi_r|h_{1k}|^2$ is expanded as
\begin{align}
&F_{\left| {h_{1k} } \right|^2 } \left( \frac{z}{\psi_r} \right) \buildrel (a) \over =  - \sum\limits_{p = 1}^\infty  {\frac{{\left( { - 1} \right)^p }}{{p!}}\left( {\frac{z}{{\psi _r \sigma _{1i}^2 }}} \right)^p \sum\limits_{i = 1}^N {\sum\limits_{t = 0}^{N - 1} {\sum\limits_{A_t } {\left( { - 1} \right)^t } } } \frac{{1 + \left( {\sum\limits_{j \in A_t } {\frac{{\sigma _{2i}^2 }}{{\sigma _j^2 }}} } \right)\left( {\frac{{\sigma _{1i}^2 }}{{\sigma _i^2 }} + \sum\limits_{j \in A_t } {\frac{{\sigma _{1i}^2 }}{{\sigma _j^2 }}} } \right)^{p - 1} }}{{1 + \sum\limits_{j \in A_t } {\frac{{\sigma _{2i}^2 }}{{\sigma _j^2 }}} }}}  \notag\\
&\buildrel (b) \over =  - \sum\limits_{p = 1}^\infty {\frac{{\left(
{ - 1} \right)^p }}{{p!}}\left( {\frac{z}{{\psi _r \sigma _{1i}^2
}}} \right)^p \sum\limits_{i = 1}^N {\sum\limits_{t = 0}^{N - 1}
{\sum\limits_{A_t } {\left( { - 1} \right)^t } } } } \left(
{\sum\limits_{j \in A_t } {\frac{{\sigma _{2i}^2 }}{{\sigma
_j^2 }}} } \right)\sum\limits_{k = 1}^{p - 1} {\binom{p-1}{k}\left( {\frac{{\sigma _{1i}^2 }}{{\sigma _{2i}^2 }}} \right)} ^k \left( {1 + \sum\limits_{j \in A_t } {\frac{{\sigma _{2i}^2 }}{{\sigma _j^2 }}} } \right)^{k - 1}  \notag\\
&\buildrel (c) \over = -\sum\limits_{p = 1}^\infty {\frac{{\left( {
- 1} \right)^p }}{{p!}}\left( {\frac{z}{{\psi _r \sigma _{1i}^2 }}}
\right)^p \sum\limits_{i = 1}^N {\sum\limits_{t = 0}^{N - 1}
{\sum\limits_{A_t } {\left( { - 1} \right)^t } } } } \sum\limits_{k
= 1}^{p - 1} {\binom{p-1}{k}\left( {\frac{{\sigma _{1i}^2 }}{{\sigma
_{2i}^2 }}} \right)} ^k \sum\limits_{q = 0}^{k -
1} {\binom{k-1}{q}} \left( {\sum\limits_{j \in A_t } {\frac{{\sigma _{2i}^2 }}{{\sigma _j^2 }}} } \right)^{q + 1}  \notag\\
&\buildrel (d) \over \approx -\frac{{\left( { - 1} \right)^N
}}{{N!}}\left( {\frac{z}{{\psi _r \sigma _{1i}^2 }}} \right)^N
\sum\limits_{i = 1}^N {\sum\limits_{t = 0}^{N - 1} {\sum\limits_{A_t
} {\left( { - 1} \right)^t } } } \left( {\sum\limits_{j \in A_t }
{\frac{{\sigma _{1i}^2 }}{{\sigma _j^2 }}} } \right)^{N - 1}
\end{align}
where (a) is satisfied by the Macraulian series
$\exp\left(x\right)=\sum\limits_{p=0}^{\infty}x^p/p!$ and the first
equation in Lemma~3; (b) is fulfilled by the binomial expansion of
$\big[ {\frac{{\sigma _{1i}^2 }}{{\sigma _i^2 }} + \sum\limits_{j
\in A_t } {\frac{{\sigma _{1i}^2 }}{{\sigma _j^2 }}} } \big]^{p - 1}
= \bigg[ {1 + \frac{{\sigma _{1i}^2 }}{{\sigma _{2i}^2 }}\big( {1 +
\sum\limits_{j \in A_t } {\frac{{\sigma _{2i}^2 }}{{\sigma _j^2 }}}
} \big)} \bigg]^{p - 1} $ and the second equation in Lemma~3; (c) is
achieved by the binomial expansion of $\big( {1 + \sum\limits_{j \in
A_t } {\frac{{\sigma _{2i}^2 }}{{\sigma _j^2 }}} } \big)^{k - 1} $;
(d) is obtained also by the second equation in Lemma~3, and ignoring
the high order infinitesimal.

The CDF of $\psi_h|h_{2k}|^2$ can be obtained similarly, thus the
asymptotic SER of $S_1$ with perfect CSI can be achieved by
\eqref{Eq:E1}, \eqref{Eq:SER}, and $\int_0^\infty {t^{2N} } \exp
\left( -{t^2}/{2}\right)dt={2^{\left(N-{1}/{2}\right)}}\Gamma
\left({1}/{2} + N\right)$ in \cite[eq.~(3.326.2)]{Gradshteyn94},
where $\Gamma\left(\cdot\right)$ is the Gamma
function\cite{Abramowitz}. Therefore, the diversity order is $N$.

If CSI is outdated, applying Macraulian series
$\exp\left(x\right)\approx1+x$, the CDF of $\psi_r|\hat h_{1k}|^2$
can be written as
\begin{align}
F_{{\left| {h_{1k} } \right|^2}} \left( \frac{z}{\psi_r} \right)
\approx  \sum\limits_{i = 1}^N {\sum\limits_{t = 0}^{N - 1}
{\sum\limits_{A_t } {\left( { - 1} \right)^t } } } \frac{{1 + \zeta
_1 }}{{1 + \sum\limits_{l \in A_t } {\frac{{\sigma _{2i}^2
}}{{\sigma _l^2 }}} }}\frac{z}{{\psi _r \sigma _{1i}^2 }}
\end{align}
where the high order infinitesimal is ignored.

The CDF of $\psi_h|h_{2k}|^2$ can be obtained similarly, thus the
asymptotic SER of $S_1$ with outdated CSI can be achieved by
\eqref{Eq:E1}, \eqref{Eq:SER}, and $\int_0^\infty {t^{2N} } \exp
\left( -{t^2}/{2}\right)dt={2^{\left(N-{1}/{2}\right)}}\Gamma
\left({1}/{2} + N\right)$ in \cite[eq. (3.326.2)]{Gradshteyn94}.
Therefore, the diversity order is $1$.

The asymptotic expressions of $S_2$ can be verified similarly.

It is noted that although the analytical expression in Proposition~2
is the lower bound, it matches tightly with the exact result,
especially in high SNR. Furthermore, diversity order reflects the
behavior of SER \emph{in high SNR}. Therefore, similar to the
previous research \cite{Song2011}, the diversity analysis, obtained
by the analytical expression, is accurate.

Another alternative method to analyze the diversity is achieved by
the SNR bounds. The end-to-end instantaneous SNR is upper bounded by
$\min \{\psi_r|h_{ji}|^2,\psi_h|h_{\bar j i}|^2\}$, and lower
bounded by $\frac{1}{2}\min \{\psi_s|h_{ji}|^2,\psi_h|h_{\bar j
i}|^2\}$. Similar to the previous analysis, the diversity order can
also be obtained.

\section*{Appendix F: Proof of Proposition~3}

For ease of analysis, the performance of diversity order is obtained
under the symmetric network, i.e., $\sigma_{ji}^2=1$ and
$\rho_{ji}=\rho$.

According to Lemma~1, we have the asymptotic performance
\begin{align}
\gamma_i=\min \big(\gamma_{1i},\gamma_{2i}\big)\approx \psi_h\min
\big(|h_{1i}|^2,|h_{2i}|^2\big).
\end{align}

Therefore, $\gamma_i$ follows the exponential distribution, i.e.,
\begin{align}
f_{\gamma_i}\big(x\big)=\frac{1}{\Gamma}\exp\big(-\frac{x}{\Gamma}\big)
\end{align}
where $ \Gamma=\psi_h/2$.

Denoting $\hat \gamma_i$ to be the outdated version of $\gamma_i$,
the received SNR by maximal-ratio combining the best $K$ relays is
\begin{align}\label{Eq:F1}
\gamma=\sum\limits_{i=0}^{N}T(\hat \gamma_{i})
\end{align}
where $T(\hat \gamma_{i})$ indicates whether $R_i$ is selected or
not, according to the RS scheme \eqref{Eq:outdatedRS}, i.e.,
\begin{align}
T\left( {\hat \gamma _i } \right) = \left\{ {\begin{array}{*{20}c}
   {0,\hat \gamma _i  < \hat \gamma ^{\left( K \right)} ;} \hfill  \\
   {\gamma _i ,\hat \gamma _i  \ge \hat \gamma ^{\left( K \right)} .} \hfill  \\
\end{array}} \right.
\end{align}
where $\hat \gamma^{\left(K\right)}$ represents the $K$th largest
value among $\left\{\hat \gamma_i | i=1,\ldots,N\right\}$.

Following the analysis of \cite{Dong2012}, the moment generating
function~(MGF) of $\gamma$ in \eqref{Eq:F1} is obtained by dividing
the total probability into $N$ disjoint events that $\hat
\gamma_{i}=\hat \gamma^{\left(K\right)}, i=1,\ldots,N$. During each
event, there are $\big(K-1\big)$ relays whose SNR is larger than
$\hat \gamma_{i}$, and other $\big(N-K\big)$ relays' SNR is smaller
than $\hat \gamma_{i}$, thus there are $\binom{N-1}{K-1}$
possibilities.

After some manipulation, the MGF of $\gamma$ in \eqref{Eq:F1} for
the symmetric network is \cite[eq. (15)]{Dong2012}
\begin{align}
&\Phi _\gamma  \left( s \right) = N\binom{N-1}{K-1}\sum\limits_{n =
0}^{N - K} {\sum\limits_{q = 0}^{N - K - n} {\sum\limits_{m = 0}^n
{\left( {\frac{4}{ - s\psi _h }} \right)} } } ^{N - n}
\notag\\
&\hspace{40mm}\times\frac{{\left( {N - K} \right)!\left( { - 1}
\right)^{q + m} }}{{\left( {N - K - n - q} \right)!m!q!\left( {n -
m} \right)!}}\frac{1}{N - n + m - m\rho ^2 }.
\end{align}

Therefore, the diversity order is $K$, according to
\cite[prop.~3]{Zheng2003}.

\newpage

\end{document}